\documentclass{article}

\usepackage{arxiv}

\usepackage[utf8]{inputenc} 
\usepackage[T1]{fontenc}    
\usepackage{hyperref}       
\usepackage{url}            
\usepackage{booktabs}       
\usepackage{amsfonts}       
\usepackage{microtype}      
\usepackage{lipsum}
\usepackage{graphicx}
\graphicspath{ {./images/} }

\usepackage{subcaption} 
\usepackage{xspace}
\usepackage{amsmath}
\usepackage{siunitx}
\usepackage{multirow}
\usepackage{adjustbox}

\title{Clapeyron Neural Networks for Single-Species Vapor-Liquid Equilibria}

\author{
	Jan Pav\v{s}ek \textsuperscript{1}, \quad Alexander Mitsos \textsuperscript{1,2,3}, \quad Elvis J. Sim \textsuperscript{1}, \quad Jan G. Rittig \textsuperscript{1*} \vspace*{2mm}\\
	\textsuperscript{1}{Process Systems Engineering (AVT.SVT), RWTH Aachen University} \\
	\textsuperscript{2}{JARA Center for Simulation and Data Science (CSD)}\\
	\textsuperscript{3}{Institute of Climate and Energy Systems ICE-1: Energy Systems Engineering, Forschungszentrum Jülich GmbH}\\
	\textsuperscript{*}{Corresponding author, \texttt{jan.rittig@rwth-aachen.de}}\\
}

\begin{document}

	\maketitle
	
	\begin{abstract}
		
		\noindent Machine learning (ML) approaches have shown promising results for predicting molecular properties relevant for chemical process design.
		However, they are often limited by scarce experimental property data and lack thermodynamic consistency. 
		As such, thermodynamics-informed ML, i.e., incorporating thermodynamic relations into the loss function as regularization term for training, has been proposed. 
		We herein transfer the concept of thermodynamics-informed graph neural networks (GNNs) from the Gibbs-Duhem to the Clapeyron equation, predicting several pure component properties in a multi-task manner, namely: vapor pressure, liquid molar volume, vapor molar volume and enthalpy of vaporization.
		We find improved prediction accuracy of the Clapeyron-GNN compared to the single-task learning setting, and improved approximation of the Clapeyron equation compared to the purely data-driven multi-task learning setting.
		In fact, we observe the largest improvement in prediction accuracy for the properties with the lowest availability of data, making our model promising for practical application in data scarce scenarios of chemical engineering practice. 
		
	\end{abstract}

	\section{Introduction}
	
	\noindent Process design requires information on multiple thermodynamic properties for the species included in the process, such as density, enthalpy and vapor pressure.
	Here, molecular ML has shown to provide accurate predictions for a variety of such properties, including both pure component properties, e.g., density~\cite{winter2025understanding} and phase transition enthalpy~\cite{leenhouts2025thermodynamics_gnn}, and mixtures, especially for activity coefficients~\cite{qin2023solvgnn,rittig2024thermodynamics_consistent,specht2024hanna,medina2026graph_margules,wahyudi2026deepthermomix}.
	
	Recently, research has particularly focused on advancing molecular ML by incorporating thermodynamic relations into the model architecture and training, in the form of hybrid and physics-informed ML, see overview in~\cite{rittig2025molecular_perspective}.
	Thermodynamics-informed approaches are particularly promising, as they do not rely on semi-empirical thermodynamic models, which introduce corresponding modeling limitations and assumptions; rather, they use differential equations to thermodynamic potentials.
	These thermodynamics-enriched ML approaches reduce the data required for training and enhance - in some cases, even guarantee - thermodynamic consistency of the predictions~\cite{rosenberger2022thermoconsistent, rittig2024thermodynamics_consistent,specht2024hanna}, making them highly promising for chemical process design applications.
	However, most thermo\-dynamics-informed ML models focus on predicting a single property of interest.
	
	We transfer the concept of thermodynamics-informed ML to single-species vapor-liquid equilibria prediction and extend it with multi-task training.
	That is, we train a GNN to predict vapor and liquid molar volume, vapor pressure and enthalpy of vaporization of a single-species system at vapor-liquid equilibrium. 
	We jointly train on all four properties in a multi-task learning setting.
	We use the Clapeyron equation as physics regularization~\cite{raissi2019physics} in the training loss function to enhance physical consistency of the network predictions, i.e., following a thermodynamics-informed approach.
	Notably in preliminary studies, we also tested a thermodynamics-consistent approach by directly embedding the Clapeyron equation into the GNN architecture.
	This, however, resulted in significantly lower prediction accuracy, which we attribute to the imbalanced dataset of the four independently measured properties making model training challenging, also cf.~\cite{alam2026equinet}. 
	The Clapeyron-informed approach is more flexible here — albeit at the expense of guaranteed consistency — as the Clapeyron equation acts as a soft constraint in the training loss and can be adapted using a weighting factor.
	We refer to our architecture as Clapeyron-GNN.
	
	A comparable approach has recently been published by Park et al.~\cite{park2025vleGNN}, who directly predict single-species vapor-liquid equilibria with a GNN. 
	However, they not only provide the graph structure of the molecules as input but also additional graph level features such as the acentric factor, which might not necessarily be available when making predictions for new molecules.
	Additionally, Kochi et al.~\cite{kochi2025thermodynamics} employ the Clausius-Clapeyron equation for physics-informed vapor pressure prediction, using \textit{chemprop} as graph encoder~\cite{heid2023chemprop}, but only predict vapor pressure and dynamic viscosity. 
	With Clapeyron-GNN, we train an architecture that predicts all four properties, vapor pressure, liquid and vapor molar volumes and enthalpy of vaporization, in a multi-task learning setting and relies purely on the molecular structure and temperature, and thereby enables easier prediction for new molecules.
	We benchmark the predictive performance of Clapeyron-GNN against a purely data-driven multi-task GNN and against four individual GNNs predicting all four properties in a single-task learning setting.

	\section{Clapeyron Graph Neural Network}
	
	\begin{figure}[h]  
		\centering
		\includegraphics[trim = 0 0 0 0, clip ,width=\textwidth]{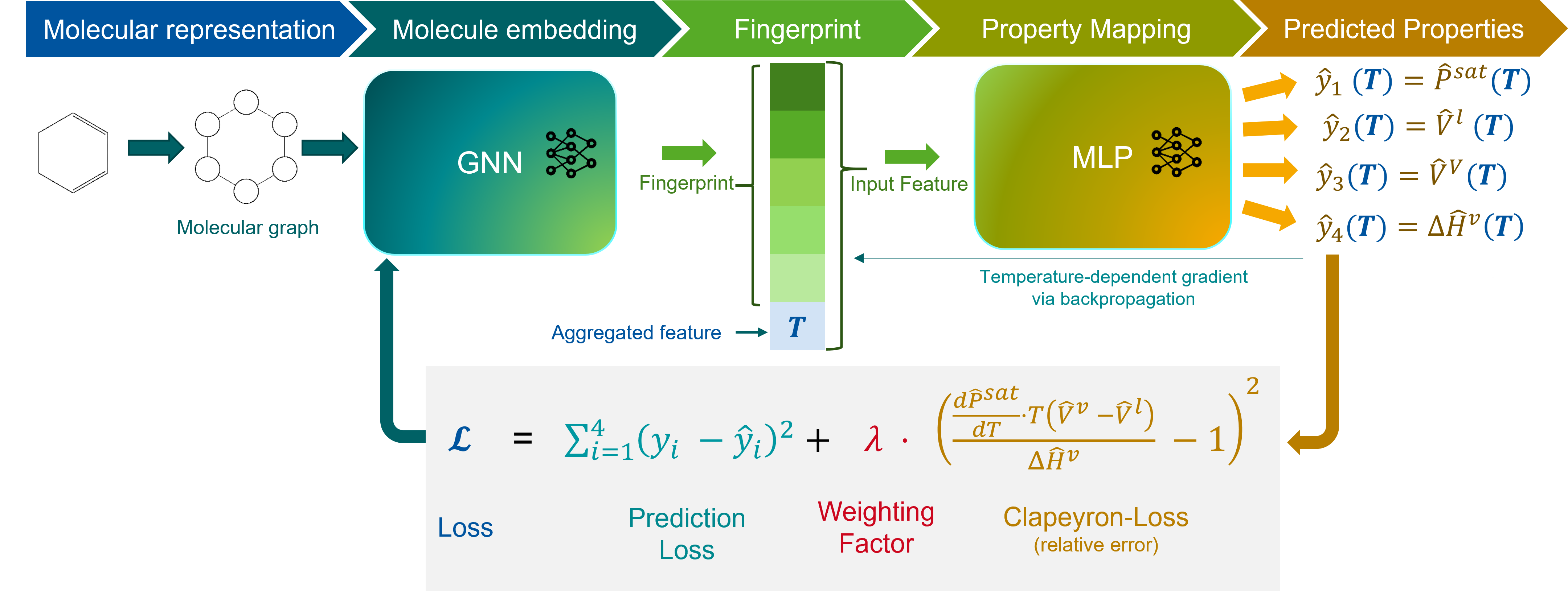}  
		\caption{Schematic illustration of Clapeyron-GNN}
		\label{fig:arch_Clap_GNN}
	\end{figure}
	
	\noindent The Clapeyron-GNN architecture is based on our previously employed GNNs~\cite{schweidtmann2020graph}. 
	That is, we map molecular graphs via graph convolutional layers and a multi-layer perceptron to molecular properties. 
	The prediction targets are the temperature-dependent vapor and liquid molar volume, the vapor pressure and the enthalpy of vaporization (see Figure \ref{fig:arch_Clap_GNN}). 
	Temperature is considered as additional input by concatenation with the molecular fingerprint.
	
	We incorporate the Clapeyron equation in the GNN model training.
	The Clapeyron equation is an exact equation derived from the equality of Gibbs free energy in the vapor and liquid phase; it provides a relationship between the four target properties and the temperature:
	
	\begin{equation} \label{eq:clapeyron_eq}
		\frac{dp^{sat}}{dT} = \frac{\Delta H_{V}}{T (V^{V} - V^{L})}
	\end{equation}
	where $p^{sat}$ is the vapor pressure, $\Delta H_{V}$ is the enthalpy of vaporization, $V^{V}$ is the vapor molar volume and $V^{L}$ is the liquid molar volume. 
	All four properties depend on the temperature $T$.
	We reformulate the Clapeyron equation to a Clapeyron error, which measures relative deviation of the network predictions from the nearest point fulfilling the Clapeyron equation:
	
	\begin{equation} \label{eq:clapeyron_error}
		\mathcal{L}_{\text{Clapeyron}} = (\frac{\frac{d\hat{p}^{sat}}{dT} T (\hat{V}^{V} - \hat{V}^{L})}{\hat{\Delta H_{V}}} - 1)^{2}.
	\end{equation}
	
	Similarly to our previous thermodynamics-informed approach~\cite{rittig2023gibbs_informed}, we extend the training loss by adding the Clapeyron error as additional regularization to the prediction error using a weighting factor $\lambda$, see Figure~\ref{fig:arch_Clap_GNN}.
	The gradient of the vapor pressure with respect to the temperature is obtained during training from backpropagation, preserving end-to-end learning.
	In contrast to the prediction loss, which can only be evaluated for each target property if data exists for the particular property, the Clapeyron regularization provides a loss signal for all four properties, even for temperature points for which only data on a subset of the target properties is available.
	This is also a major practical difference between employing a thermodynamics-informed and a thermodynamics-consistent approach. 
	As thermodynamics-consistent models embed the thermodynamics equations directly in the output head of the model, they rely on the same number of loss signals as purely data-driven models. 
	We performed first investigations of employing a thermodynamics-consistent approach, ie., directly embedding the exact Clapeyron equation into the output head, which resulted in non-converging training and lower prediction performance than purely data-driven models.
	Hence, in the present work we employ the thermodynamics-\emph{informed} Clapeyron-GNN.
	
	\section{Case Study}
	
	\noindent
	
	\subsection{Dataset}\label{subsec:data}
	
	\begin{figure}[bt]
		\centering
		
		\begin{subfigure}[t]{0.47\textwidth}
			\centering
			\includegraphics[width=\linewidth]{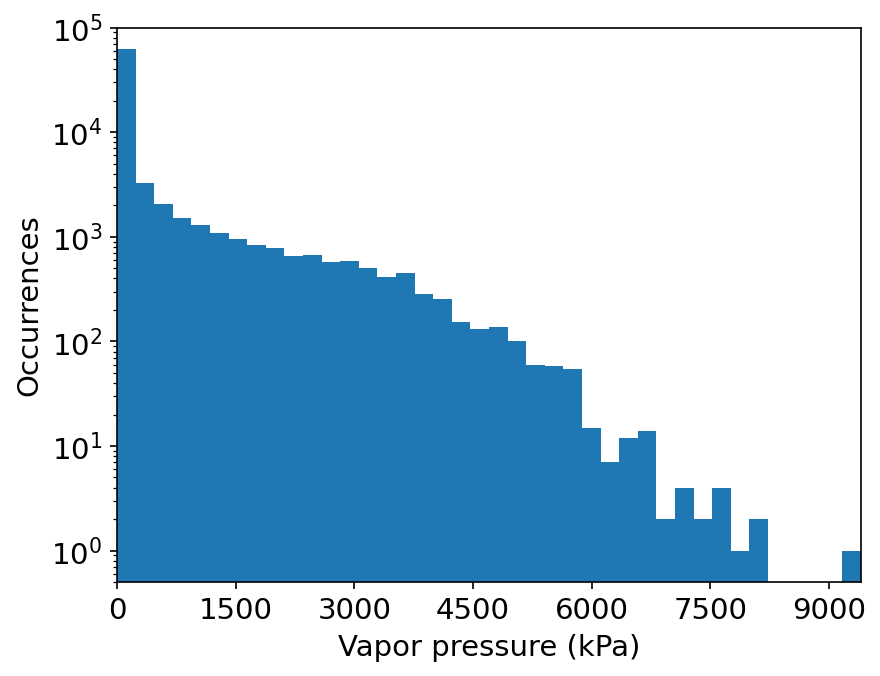}
			\caption{$p^{sat}\text{(T)}$}
		\end{subfigure}
		\hfill
		\begin{subfigure}[t]{0.47\textwidth}
			\centering
			\includegraphics[width=\linewidth]{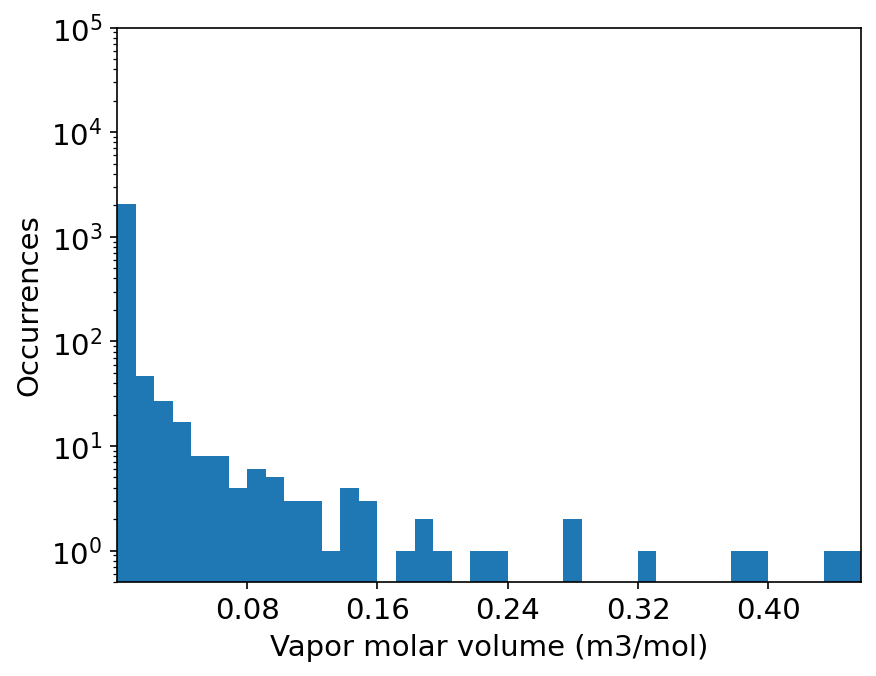}
			\caption{$V^{V}\text{(T)}$}
		\end{subfigure}
		
		\par\medskip
		
		\begin{subfigure}[t]{0.47\textwidth}
			\centering
			\includegraphics[width=\linewidth]{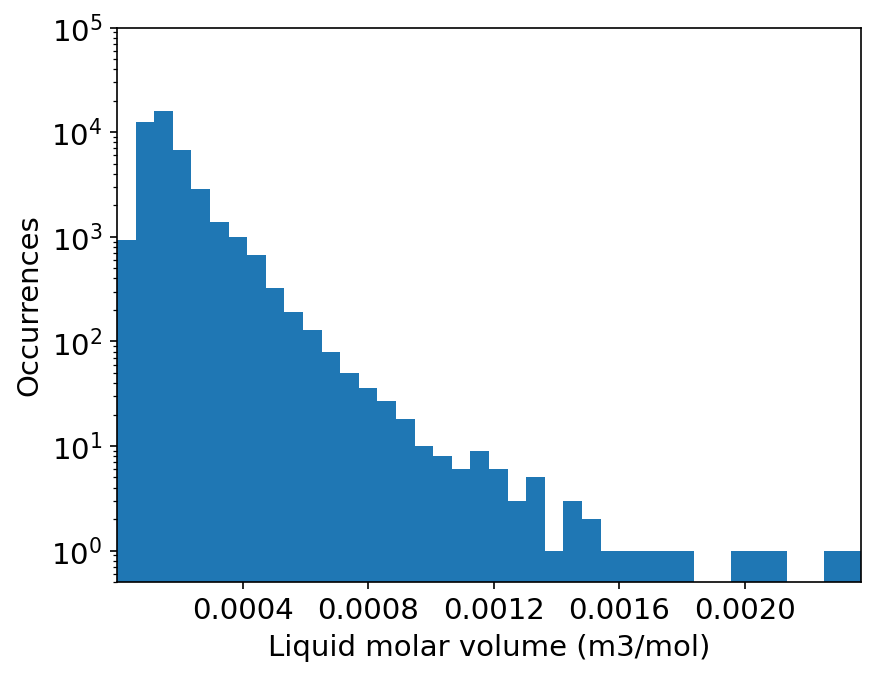}
			\caption{$V^{L}\text{(T)}$}
		\end{subfigure}
		\hfill
		\begin{subfigure}[t]{0.47\textwidth}
			\centering
			\includegraphics[width=\linewidth]{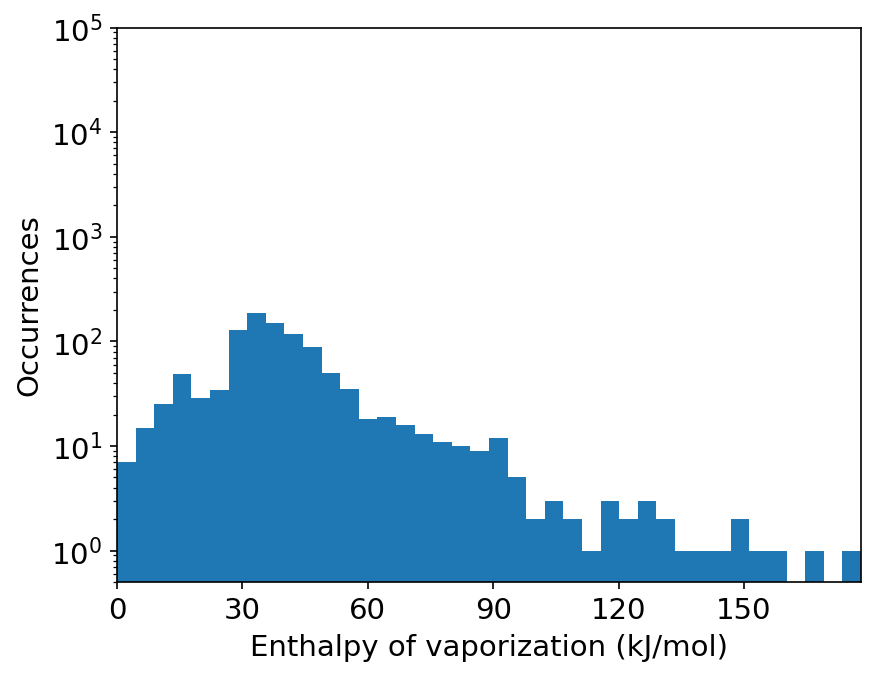}
			\caption{$\Delta H_{V}\text{(T)}$}
		\end{subfigure}
		
		\caption{Histograms of data distributions for four target properties and 40 bins for numeric values across all properties}
		\label{fig:data_distribution}
	\end{figure}
	
	\noindent We extract experimental data for the vapor and liquid molar volume, the vapor pressure and the enthalpy of vaporization from the NIST ThermoData Engine~\cite{nist_tde}. 
	Notably, we manually removed 10 outliers, whose numeric values deviated by at least an order of magnitude from the rest of the dataset.
	Our dataset contains 879 molecules at varying temperatures, ranging from 56.75~K to 1021~K, with 102,121 data-points in total. 
	The molecules are organic compounds, including amines, esters, alcohols, carboxylic acids, ketones, phenols, amines, nitro compounds and amides, and range in molecular weight from 27 to 493$\frac{\text{g}}{\text{mol}}$.
	The distribution of data points across different molecules and the four different properties is highly uneven (see Figure \ref{fig:data_distribution}). 
	Specifically, the amount of data points varies from a single point to more than a thousand for different molecules.
	Considering the properties, our data sets contains 78,840 data points for the vapor pressure, 43,056 for liquid molar volume, 2,206 for vapor molar volume and 1,057 for enthalpy of vaporization. 
	In fact, there is abundant training data across the temperature domain for the vapor pressure and the liquid molar volume, whereas for the vapor molar volume and the enthalpy of vaporization, most molecules have either a single or no data point at all.
	Therefore, learning the temperature dependency for these two properties is inherently more difficult.
	Not only does the number of available datapoints vary largely between properties, but also within one property, the distribution of numeric values is skewed (see Figure \ref{fig:data_distribution}).
	To account for this, all four properties are predicted on a logarithmic scale; temperature values are linearly normalized to range 0-1. 
	All error metrics are therefore reported on the logarithmic scale.

	\subsection{Prediction Scenario \& Benchmark}
	
	\noindent We evaluate the prediction performance for extrapolating to new molecules. 
	Hence, we randomly select all 80\% of the molecules with corresponding data points for the training set and keep the remaining 20\% of molecules and associated data points for model testing. 
	To evaluate the contribution of the multi-task learning setting and the physics regularization, we compare Clapeyron-GNN against two base cases: (i) a purely data driven multi-task learning setting, which we refer to as MTL-GNN, and (ii) a purely data driven single-task learning setting, where we train separate models for each property, which we refer to as STL-GNN.
	We train and evaluate each model 10 times with different seeds, and report average and standard deviation across these runs in all metrics.

	\subsection{Implementation \& Hyperparameters}\label{subsec:impl_hyp}
	
	\noindent The model is implemented using PyTorch and PyTorch Geometric~\cite{FeyLenssen2019pytorchgeom} within our \href{https://git.rwth-aachen.de/avt-svt/public/gmolprop/}{\emph{GMoLprop}} framework, which is available as open source. 
	Hyperparameters are optimized with grid search, where the optimized parameters are the batch size $\in$ \{64, 128\}, the activation function $\in$ \{LeakyReLU, SiLU\}, the fingerprint dimension $\in$ \{64, 128\} and for Clapeyron-GNN, the weighting factor of the Clapeyron loss $\in$ \{0.1, 0.5, 1\}.
	This results in a batch size of 64, a fingerprint dimension of 64 and LeakyReLU as activation function for the STL-GNN, a batch size of 64, a fingerprint dimension of 64 and LeakyReLU as activation function for the MTL-GNN, and a batch size of 64, a fingerprint dimension of 128, LeakyReLU as activation function and a weighting factor of 0.1 for the Clapeyron-GNN.
	This value of the weighting factor yields the best performance as it does not negatively impact data approximation or destabilize training, which is observed at larger values for the weighting factor, while still achieving good approximation of the Clapeyron equation.
	For the activation function, LeakyReLU results in better predictive performance than SiLU.
	While SiLU yields smooth functions with respect to temperature and lower Clapeyron errors, it significantly decreases predictive performance.
	Hence, we choose LeakyReLU as activation function.
	
	\section{Results \& Discussion}
	
	\noindent We first compare the Claperyon-GNN to the other models, and then show exemplary predictions for individual molecules.
	
	\subsection{Model comparison} \label{subsec:res_model_comp}
	
	\begin{table}[h!]
		\centering
		\caption{Performance metrics, root mean squared error (RMSE), mean absolute error (MAE), and coefficient of determination ($\text{R}^{2}$) evaluated on the logarithmic scale on the test set for the three different models and for all four properties, and Clapeyron error (see Equation \ref{eq:clapeyron_error}) evaluated on the test set. Large font values are averages and small fonts standard deviations.}
		\label{tab:perform_met}
		\begin{adjustbox}{max width=1.0\textwidth,center}
			\begin{tabular}{c|c c c|c c c|c c c}
				
				\multirow{2}{*}{Property} & \multicolumn{3}{c|}{STL-GNN} & \multicolumn{3}{c|}{MTL-GNN} & \multicolumn{3}{c}{Clapeyron-GNN}  \\
				& RMSE & MAE & $\text{R}^{2}$ & RMSE & MAE & $\text{R}^{2}$ & RMSE & MAE & $\text{R}^{2}$ \\
				\hline
				
				$p^{sat}\text{(T)}$ & \multirow{2}{*}{\textbf{0.27} {\tiny $\pm$ 0.019}} & \multirow{2}{*}{\textbf{0.14} {\tiny $\pm$ 0.0074}} & \multirow{2}{*}{\textbf{0.97} {\tiny $\pm$ 0.0040}} & \multirow{2}{*}{\textbf{0.26
					} {\tiny $\pm$ 0.013}} & \multirow{2}{*}{\textbf{0.14} {\tiny $\pm$ 0.0076}} & \multirow{2}{*}{\textbf{0.97} {\tiny $\pm$ 0.0025}} & \multirow{2}{*}{\textbf{0.26} {\tiny $\pm$ 0.019}} & \multirow{2}{*}{\textbf{0.14} {\tiny $\pm$ 0.0065}} & \multirow{2}{*}{\textbf{0.97} {\tiny $\pm$ 0.0040}} \\
				{\small(\#78,840)} & & & & & & & & & \\
				\hline
				$V^{V}\text{(T)}$& \multirow{2}{*}{0.31 {\tiny $\pm$ 0.028}} & \multirow{2}{*}{0.18 {\tiny $\pm$ 0.022}} & \multirow{2}{*}{0.84 {\tiny $\pm$ 0.028}} & \multirow{2}{*}{\textbf{0.17} {\tiny $\pm$ 0.022}} & \multirow{2}{*}{\textbf{0.13} {\tiny $\pm$ 0.021}} & \multirow{2}{*}{\textbf{0.95} {\tiny $\pm$ 0.013}} & \multirow{2}{*}{\textbf{0.18} {\tiny $\pm$ 0.019}} & \multirow{2}{*}{\textbf{0.14} {\tiny $\pm$ 0.016}} & \multirow{2}{*}{\textbf{0.95} {\tiny $\pm$ 0.011}} \\
				{\small(\#2,206)} & & & & & & & & & \\
				\hline
				$V^{L}\text{(T)}$ & \multirow{2}{*}{\textbf{0.048} {\tiny $\pm$ 0.0057}} & \multirow{2}{*}{\textbf{0.020} {\tiny $\pm$ 0.0012}} & \multirow{2}{*}{\textbf{0.95} {\tiny $\pm$ 0.013}} & \multirow{2}{*}{\textbf{0.048} {\tiny $\pm$ 0.0029}} & \multirow{2}{*}{0.030 {\tiny $\pm$ 0.0022}} & \multirow{2}{*}{\textbf{0.95} {\tiny $\pm$ 0.0064}} & \multirow{2}{*}{\textbf{0.048} {\tiny $\pm$ 0.0050}} & \multirow{2}{*}{0.029 {\tiny $\pm$ 0.0046}} & \multirow{2}{*}{\textbf{0.95} {\tiny $\pm$ 0.012}} \\
				{\small(\#43,056)} & & & & & & & & & \\
				\hline
				$\Delta H_{V}\text{(T)}$& \multirow{2}{*}{0.15 {\tiny $\pm$ 0.024}} & \multirow{2}{*}{0.099 {\tiny $\pm$ 0.017}} & \multirow{2}{*}{0.71 {\tiny $\pm$ 0.096}} & \multirow{2}{*}{\textbf{0.11} {\tiny $\pm$ 0.016}} & \multirow{2}{*}{\textbf{0.083} {\tiny $\pm$ 0.017}} & \multirow{2}{*}{\textbf{0.85} {\tiny $\pm$ 0.046}} & \multirow{2}{*}{\textbf{0.10} {\tiny $\pm$ 0.023}} & \multirow{2}{*}{\textbf{0.075} {\tiny $\pm$ 0.018}} & \multirow{2}{*}{\textbf{0.85} {\tiny $\pm$ 0.063}} \\
				{\small(\#1,057)} & & & & & & & & & \\
				\noalign{\hrule height 1.5pt}
				$\mathcal{L}_{\text{Clapeyron}}$& \multicolumn{3}{c|}{0.45 {\tiny $\pm$ 0.12}} & \multicolumn{3}{c|}{0.14 {\tiny $\pm$ 0.050}} & \multicolumn{3}{c}{\textbf{0.0069} {\tiny $\pm$ 0.0051}}  \\
				\hline
				
			\end{tabular}
		\end{adjustbox}
	\end{table}
	
	\noindent The overall prediction accuracy for all four properties across the three different models, STL-GNN, MTL-GNN and Clapeyron-GNN over 10 runs is shown in Table \ref{tab:perform_met}. 
	
	Comparing the performance of the MTL-GNN to the STL-GNNs, a clear increase in prediction performance can be observed for the vapor molar volume and the enthalpy of vaporization, for which data is rather scarce, see Section~\ref{subsec:data}.
	For the liquid molar volume and the vapor pressure with relatively high data availability, the performance is comparable across the MTL-GNN and the STL-GNN. 
	The relative improvement in the prediction performance of vapor molar volume and enthalpy of vaporization by the introduction of multi-task learning is on par, i.e. the RMSE reduces from 0.31 to 0.17 and from 0.15 to 0.11, respectively. 
	As the amount of data on vapor molar volume is twice as large as on the enthalpy of vaporization, a notable difference in the performance improvement through introduction of multi-task learning might have been expected.
	However, as this is not observed, small variations in dataset size appear not to have a significant influence on the value of multi-task learning.
	Performance improvements become observable at dataset size differences across orders of magnitude in our case. 
	Hence in our case, multi-task learning introduces a clear advantage for properties where data are scarce, while for properties where data are abundant in the temperature domain, the prediction performance is not altered significantly. 
	This aligns with the findings by bin Javaid et al.~\cite{bin2025exploring_augment}, who find that MTL improves prediction accuracy only for tasks that have strong relationships among each other and limited data availability.
	Standley et al.~\cite{standley2020tasks} report that in data scarce scenarios the change in predictive performance of MTL compared to STL depends largely on the combination of tasks that are trained jointly, significantly improving the performance for some and even worsening it for others, which can depend on the relationships between tasks.
	As in the present work, all four properties are strongly related - the existence of the Clapeyron equation serves as proof for that - our results align with literature reports with respect to the effect of MTL on predictive performance.
	
	The prediction performance of the Clapeyron-GNN is on par with the high prediction performance of the MTL-GNN for all four prediction targets. 
	However, in the Clapeyron error (see Table \ref{tab:perform_met}) a significant difference can be observed between the Clapeyron-GNN and the MTL-GNN.
	The Clapeyron error is with 0.007 compared to 0.138, respectively, two orders of magnitude smaller when using the thermodynamics-informed approach. 
	Thus, for the single-species VLE prediction, Clapeyron regularization can significantly enhance the approximation of fundamental thermodynamics relations, without negatively impacting the performance in data approximation.
	The reason why it does not improve the prediction performance lies in the fact that the Clapeyron regularization provides additional information on the consistency of individual molecules.
	However, it does not give additional information on new molecules.
	As we test on unseen molecules, the prediction performance of the Clapeyron-GNN remains on the same level but rather shows improved consistency of predictions.
	Notably, also the introduction of MTL alone enhances adherence to the Clapeyron equation; for the STL-GNN, the Clapeyron error is significantly higher than for the MTL-GNN, with 0.45 compared to 0.14, respectively (see Table \ref{tab:perform_met}). 
	So interestingly, the introduction of multi-task learning also improves the approximation of the Clapeyron equation, albeit significantly less than the introduction of Clapeyron regularization.
	
	Overall, the Clapeyron-GNN achieves high prediction accuracy on the same level as MTL, while significantly better approximating fundamental thermodynamic relations.
	We thus further investigate the prediction capabilities of the Clapeyron-GNN for the best models and individual molecules in the following. \\
	
	\begin{figure}[p]
		\vspace*{-1.5cm}
		\centering
		
		\begin{subfigure}[t]{0.38\textwidth}
			\centering
			\includegraphics[width=\linewidth]{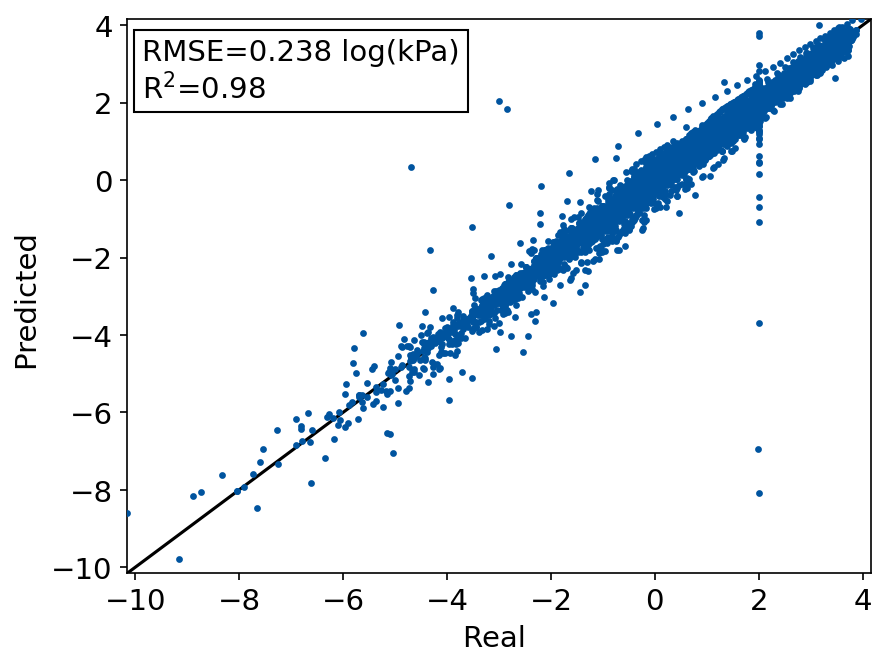}
			\caption{Vapor pressure MTL-GNN}
		\end{subfigure}
		\hfill
		\begin{subfigure}[t]{0.38\textwidth}
			\centering
			\includegraphics[width=\linewidth]{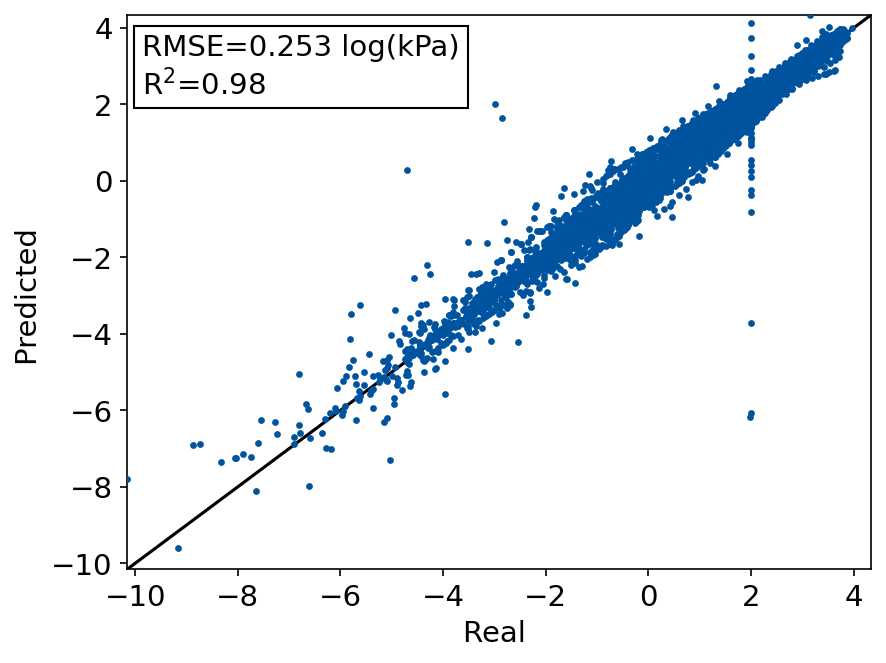}
			\caption{Vapor pressure Clapeyron-GNN}
		\end{subfigure}
		
		\par\medskip
		
		\begin{subfigure}[t]{0.38\textwidth}
			\centering
			\includegraphics[width=\linewidth]{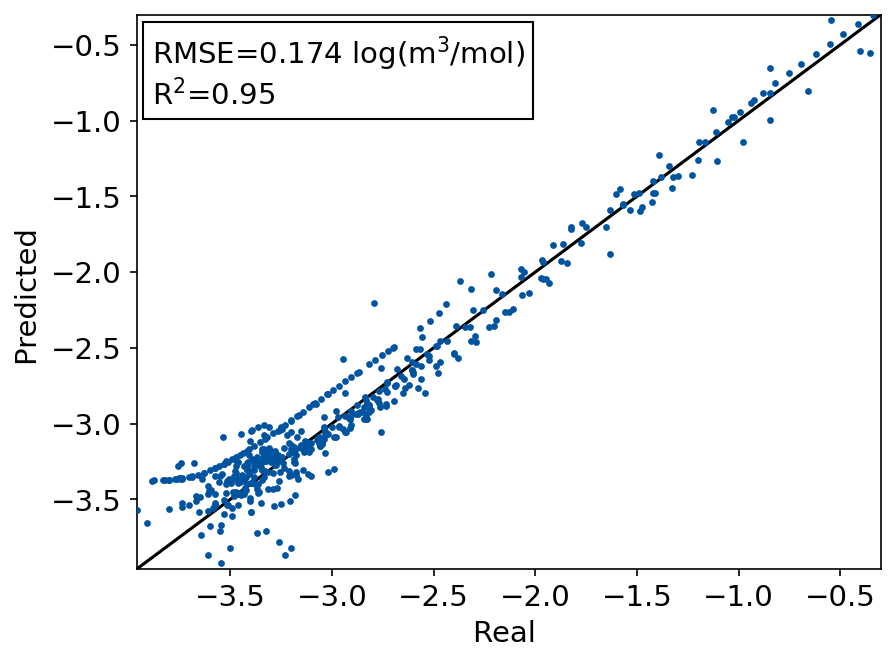}
			\caption{Vapor molar volume MTL-GNN}
		\end{subfigure}
		\hfill
		\begin{subfigure}[t]{0.38\textwidth}
			\centering
			\includegraphics[width=\linewidth]{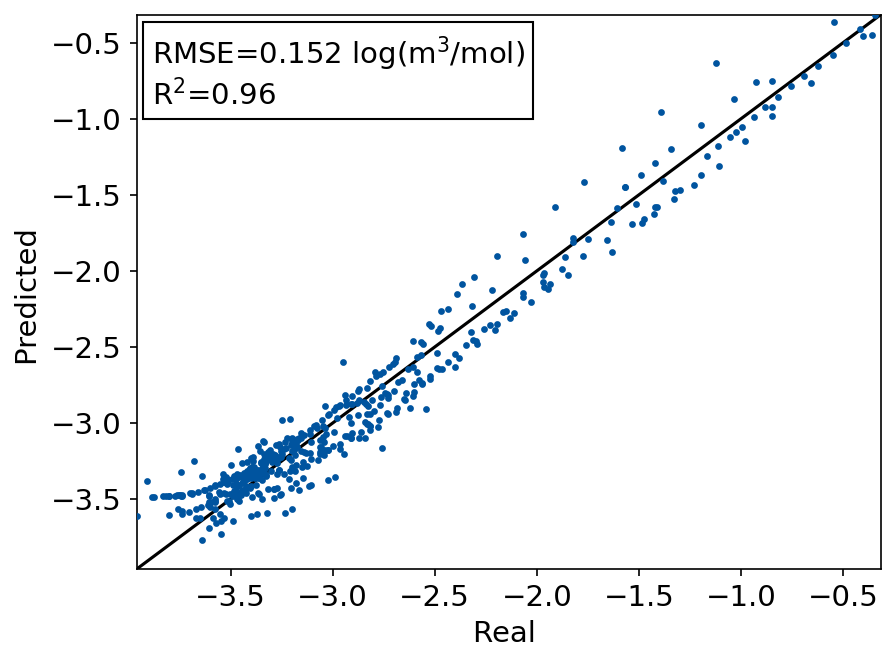}
			\caption{Vapor molar volume Clapeyron-GNN}
		\end{subfigure}
		
		\par\medskip
		
		\begin{subfigure}[t]{0.38\textwidth}
			\centering
			\includegraphics[width=\linewidth]{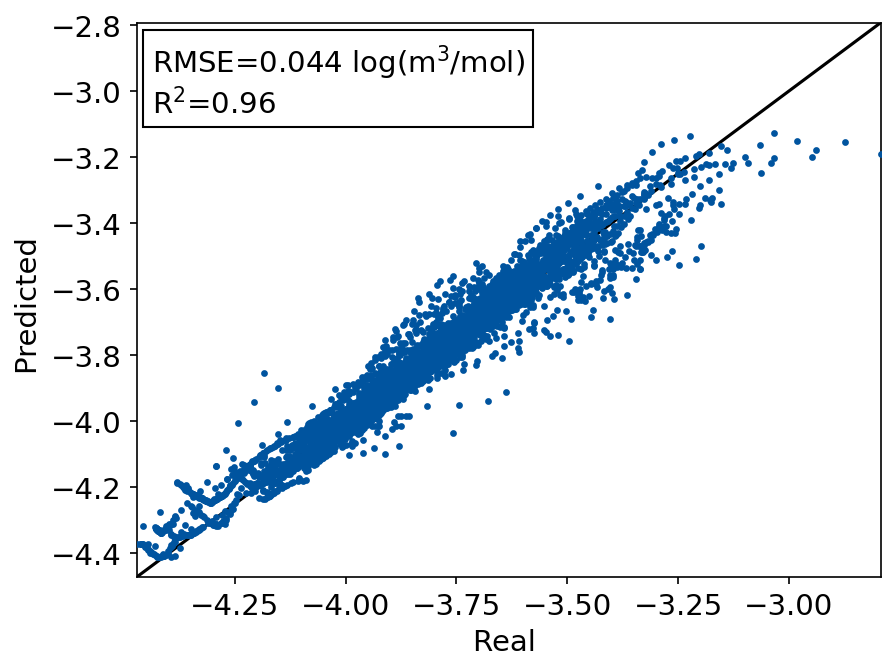}
			\caption{Liquid molar volume MTL-GNN}
		\end{subfigure}
		\hfill
		\begin{subfigure}[t]{0.38\textwidth}
			\centering
			\includegraphics[width=\linewidth]{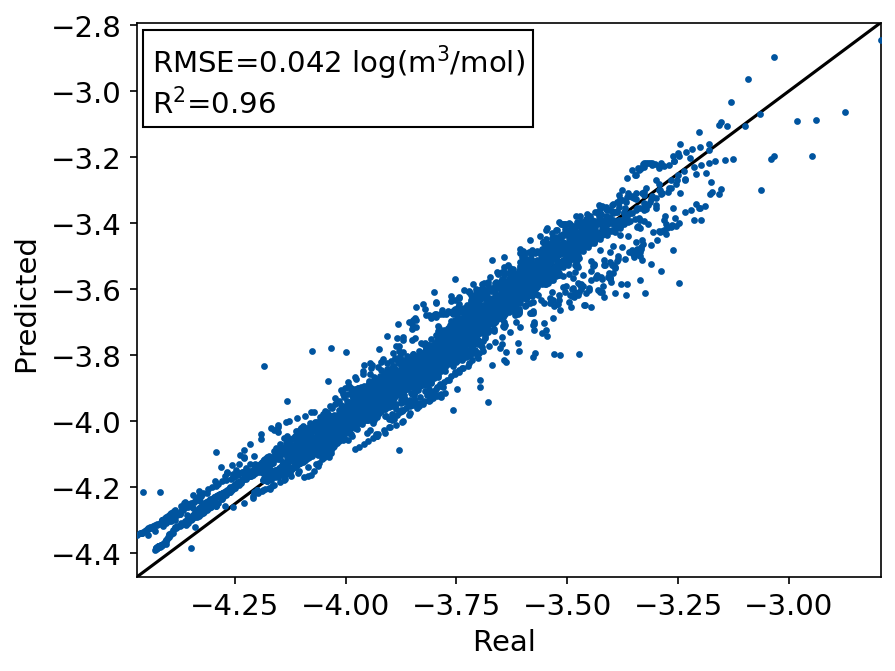}
			\caption{Liquid molar volume Clapeyron-GNN}
		\end{subfigure}
		
		\par\medskip
		
		\begin{subfigure}[t]{0.38\textwidth}
			\centering
			\includegraphics[width=\linewidth]{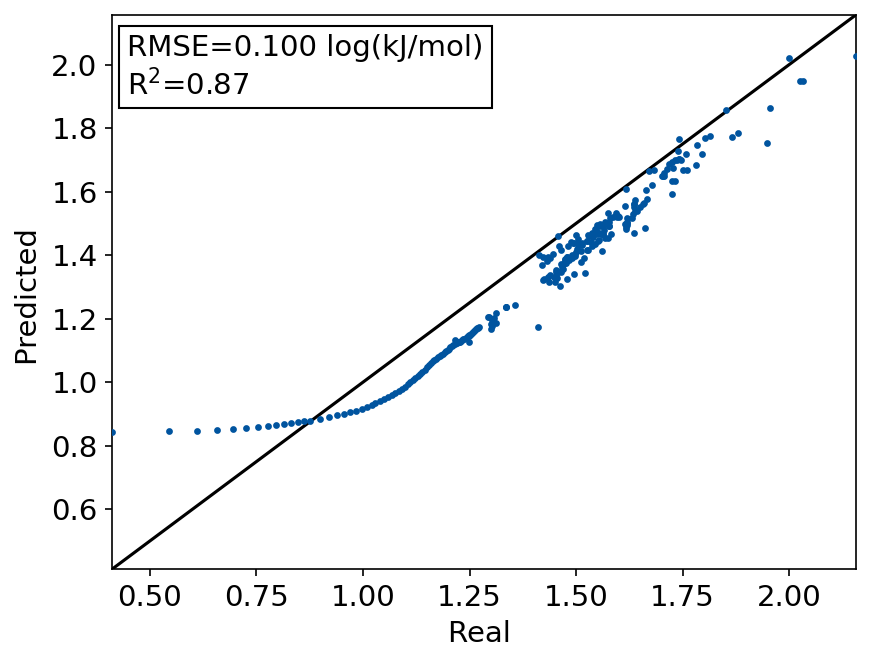}
			\caption{Enthalpy of vaporization MTL-GNN}
		\end{subfigure}
		\hfill
		\begin{subfigure}[t]{0.38\textwidth}
			\centering
			\includegraphics[width=\linewidth]{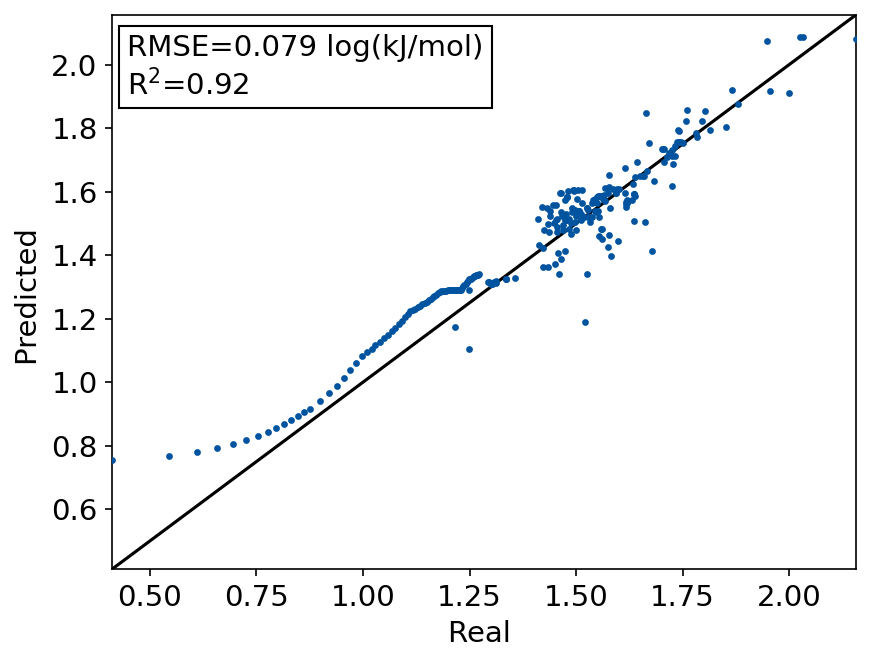}
			\caption{Enthalpy of vaporization Clapeyron-GNN}
		\end{subfigure}
		
		\caption{Parity plots of test set MTL-GNN and Clapeyron-GNN}
		\label{fig:parity}
	\end{figure}
	
	\noindent We show the parity plots of the best model out of the 10 different seeds, for all four properties for the MTL-GNN and the Clapeyron-GNN, see Figure \ref{fig:parity}.
	We do not show the STL-GNN due to low accuracy.
	Overall, the best model of the Clapeyron-GNN slightly outperforms that of the MTL-GNN in two of the four properties ($V^{V}\text{(T)}$, $\Delta H_{V}\text{(T)}$).
	In all four properties, parallel lines to the diagonal are visible both for the predictions by the Clapeyron-GNN and by the MTL-GNN.
	This indicates that for individual molecules, the predictions follow the trend in the temperature dependency but contain an off-set. 
	
	For the vapor molar volume and for the enthalpy of vaporization, see Figure~\ref{fig:parity}(c)-(d)~\&~(g)-(h), the predictions of the Clapeyron-GNN align more closely aligned with the diagonal than for the MTL-GNN.
	However, both models overestimate the vapor molar volume in low molar volume regions (numeric values under -3.5).
	For most molecules, this region is close to the critical point, where predictions are inherently more difficult. 
	
	Overall -- also for the best models of the MTL-GNN and the Clapeyron-GNN -- the predictions are on a high level and capture the temperature dependency of the properties, with minor systematic deviations in both models on the verge of the respective data domains.

	\subsection{Predictions for individual molecules}
	
	\begin{figure}[p]
		\vspace*{-0.8cm}
		\centering
		
		\begin{subfigure}[t]{0.38\textwidth}
			\centering
			\includegraphics[width=\linewidth]{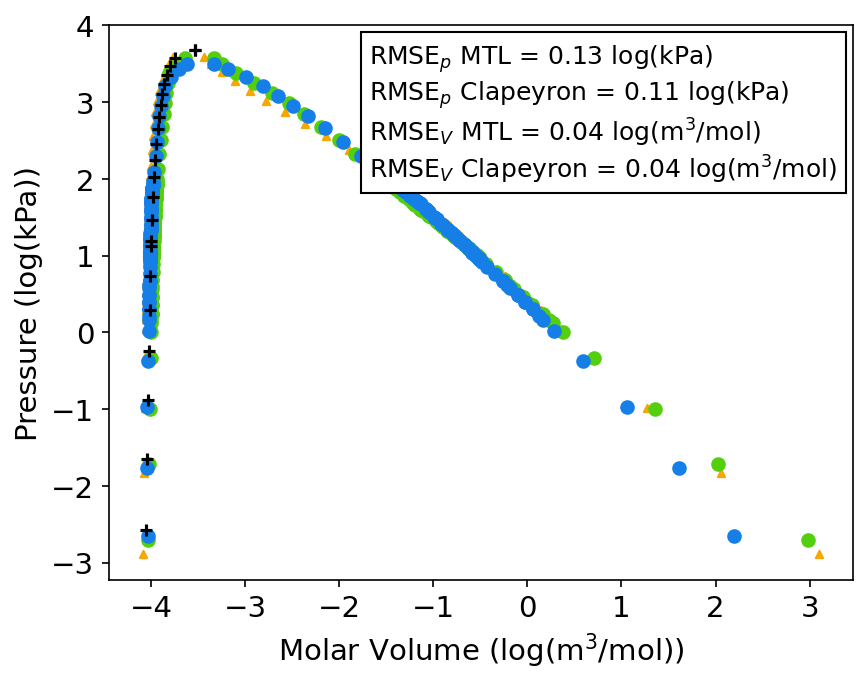}
			\caption{3-Methylthiophene}
		\end{subfigure}
		\hfill
		\begin{subfigure}[t]{0.38\textwidth}
			\centering
			\includegraphics[width=\linewidth]{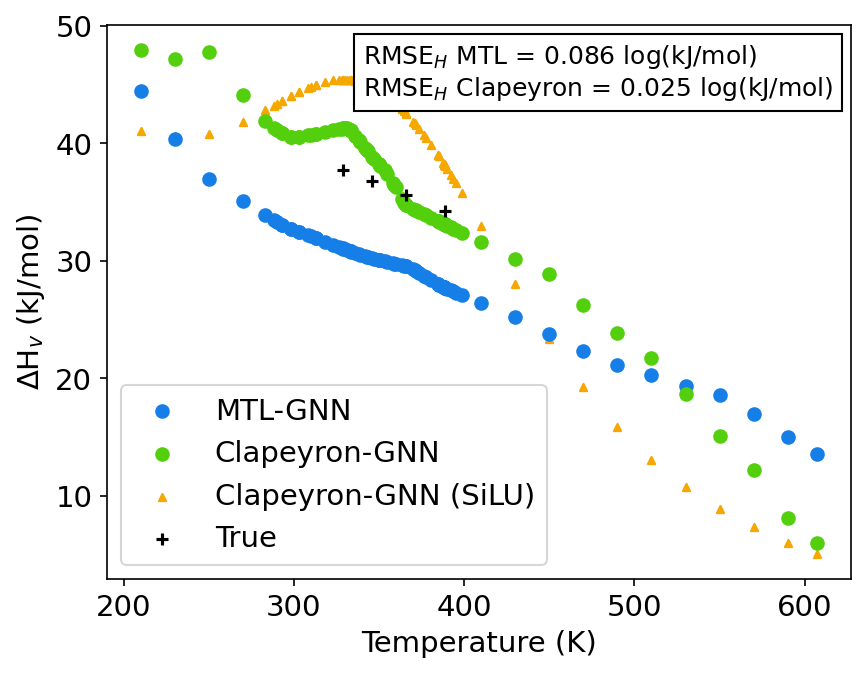}
			\caption{3-Methylthiophene}
		\end{subfigure}
		
		\par\medskip
		
		\begin{subfigure}[t]{0.38\textwidth}
			\centering
			\includegraphics[width=\linewidth]{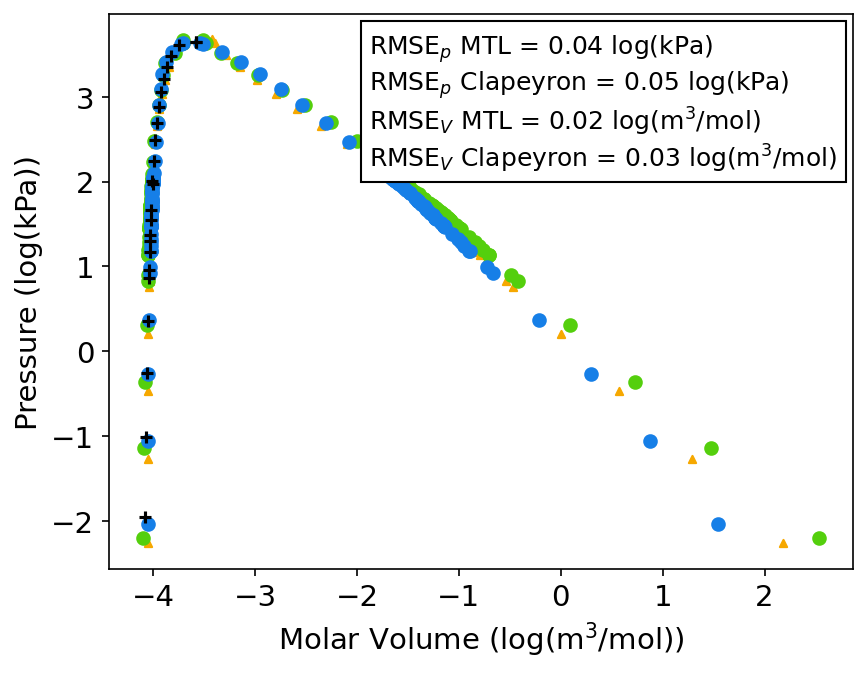}
			\caption{2-Bromopropane}
		\end{subfigure}
		\hfill
		\begin{subfigure}[t]{0.38\textwidth}
			\centering
			\includegraphics[width=\linewidth]{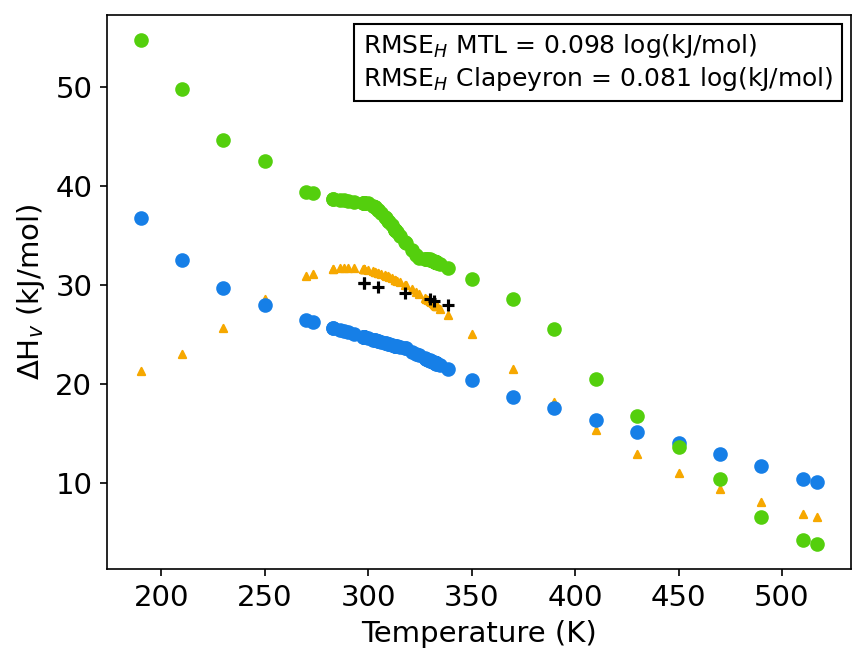}
			\caption{2-Bromopropane}
		\end{subfigure}
		
		\par\medskip
		
		\begin{subfigure}[t]{0.38\textwidth}
			\centering
			\includegraphics[width=\linewidth]{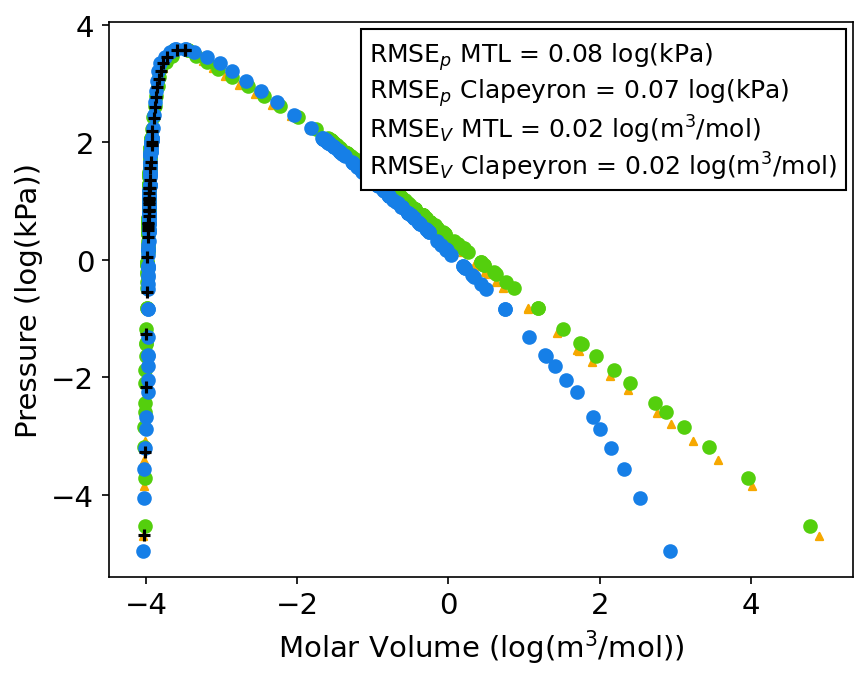}
			\caption{1-Bromobutane}
		\end{subfigure}
		\hfill
		\begin{subfigure}[t]{0.38\textwidth}
			\centering
			\includegraphics[width=\linewidth]{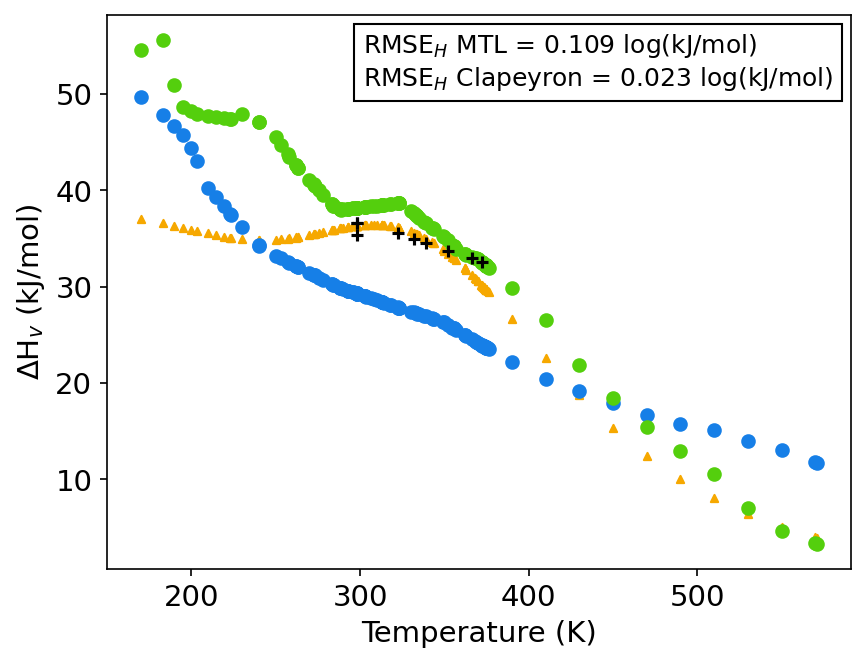}
			\caption{1-Bromobutane}
		\end{subfigure}
		
		\par\medskip
		
		\begin{subfigure}[t]{0.38\textwidth}
			\centering
			\includegraphics[width=\linewidth]{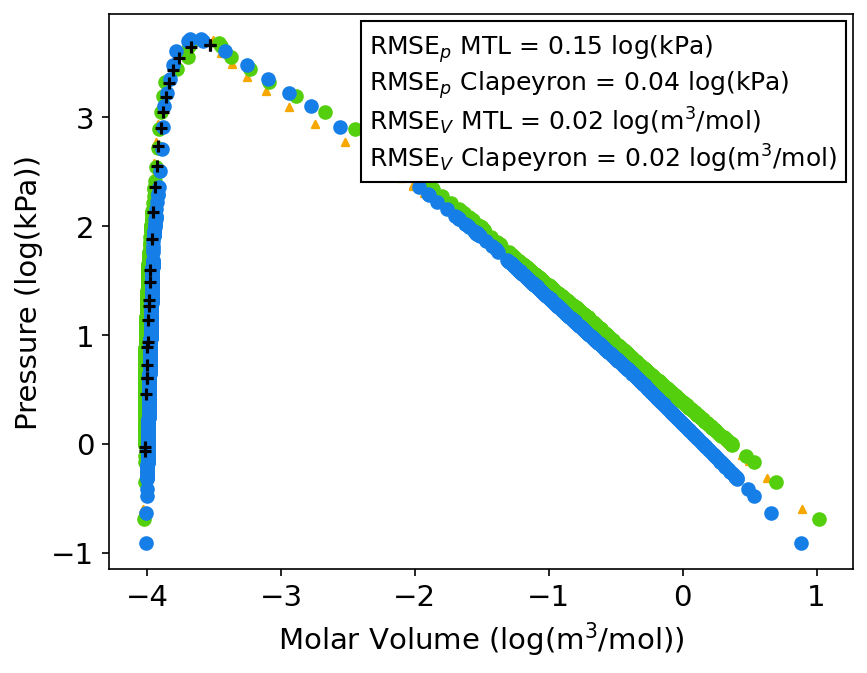}
			\caption{Piperidine}
		\end{subfigure}
		\hfill
		\begin{subfigure}[t]{0.38\textwidth}
			\centering
			\includegraphics[width=\linewidth]{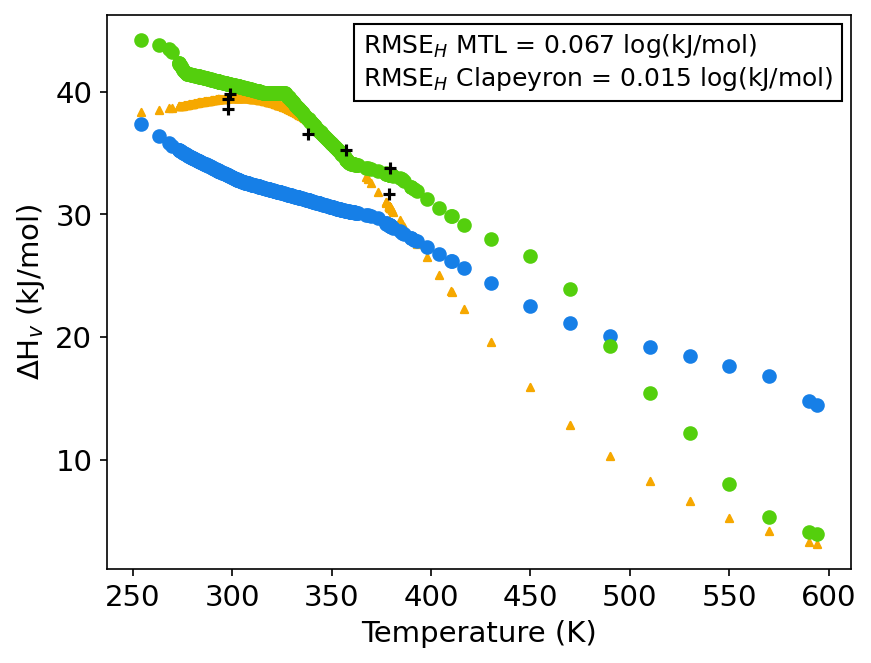}
			\caption{Piperidine}
		\end{subfigure}
		
		\caption{p(V)-plots and $\Delta$H$_{\text{v}}$(T)-plots for four exemplary molecules of the test set: experimental data in black crosses, multi-task learning in blue dots, Clapeyron-informed learning with LeakyReLU in green dots, and Clapeyron-informed learning with SiLU in orange triangles.}
		\label{fig:ind_mol_example}
	\end{figure}
	
	\noindent In Figure \ref{fig:ind_mol_example}, we show the p(V)-plot and $\Delta$H$_{\text{v}}$(T)-plot for predictions of the MTL-GNN (in blue) the Clapeyron-GNN (in green), and the Clapeyron-GNN with SiLU as activation function (in orange) for four molecules from the test set, namely 1-Bromobutane, 2-Bromopropane, 3-Methylthiophene and Piperidine.
	The four representative molecules are randomly selected from the subset of molecules in the test set that contain more than one datapoint for the enthalpy, to allow for comparison with experimental data.
	
	In the p(V)-plots, good alignment of both the Clapeyron-GNN and MTL-GNN predictions with the experimental data can be observed across orders of magnitude in the pressure and the molar volumes.
	In the lower pressure region, the vapor molar volume predictions deviate significantly between the MTL-GNN and the Clapeyron-GNN, especially for 1-Bromobutane \ref{fig:ind_mol_example}~(e).
	As in this area, there is no experimental data, it is not clear which prediction describes the true behavior more closely.
	For 1-Bromobutane and 2-Bromopropane, both the Clapeyron-GNN and the MTL-GNN approximate the critical point well.
	For 3-Methylthiophene both models underestimate the numeric values around the critical point, with the MTL-GNN underestimating more significantly, while for Piperidine the MTL-GNN overestimates the numeric values.
	Accurate predictions close to the critical point are inherently difficult, underlining the quality of the Clapeyron-GNN and MTL-GNN predictions, particularly as the models have not been trained on data from these specific molecules.
	
	For the enthalpy, a significant difference in the prediction accuracy is observable. 
	The Clapeyron-GNN approximates the data significantly better but exhibits a corner, while the predictions by the MTL-GNN have a constant off-set to the experimental data. 
	This off-set is also visible in the parity plot for enthalpy of vaporization of the MTL-GNN, where most prediction lie below the diagonal (see Section \ref{subsec:res_model_comp}).
	
	Apart from the approximation of the data, there is also a large difference in the trend of the MTL-GNN and the Clapeyron-GNN at higher temperatures.
	While the MTL-GNN predicts a linear trend with constant gradient with respect to temperature which seems close to the gradient in the experimental data, the Clapeyron-GNN predicts a steeper curve at higher temperatures, predicting significantly lower enthalpies at high temperatures than the MTL-GNN.
	These predictions are more consistent as the enthalpy of vaporization approaches zero, as the temperature approaches the critical temperature. 
	Given the fact, that there is little data on the enthalpy across the temperature domain for most molecules, learning this trend from data alone is challenging.
	This is where the strength of the Clapeyron-regularization can be observed, as it improves the consistency of predictions particularly in these areas close to the critical point where data are scarce, and the purely data-driven MTL-GNN fails to make consistent predictions.
	
	However, flaws can also be observed in the Clapeyron-GNN predictions. 
	At medium to lower temperatures corner points are visible in the Clapeyron-GNN predictions.
	This behavior is non-physical as the enthalpy of vaporization decreases in a smooth curve from its maximum at the triple point until it approaches zero at the critical point. 
	Notably, when employing SiLU as activation function, the Clapeyron-GNN yields smooth output functions, but has a local maximum in the enthalpy for all four example molecules.
	While there are corners in the prediction of the enthalpy when using LeakyReLU as activation function for the Clapeyron-GNN, it does follow the overall trend of the data better, even beyond the temperature range in the training data.
	Hence, LeakyReLU is the better choice for balancing accuracy and consistency (see Section \ref{subsec:impl_hyp}).
	As data are scarce for the enthalpy, it is likely that the loss signal during training is dominated by the Clapeyron regularization, which might introduce corner points when there is inconsistencies in the experimental data between the four properties. 
	This underlines that thermodynamics-informed models only promote but do not guarantee consistency.
	If data were sufficiently dense across the temperature domain for all four properties, the minimization of the data approximation error would make such non-physical predictions unlikely.
	However, especially when data are scarce, non-physical predictions are possible.
	
	Nonetheless, Clapeyron-GNN shows highly promising performance, enabling good approximation of Clapeyron-consistent VLE calculation for single-species systems, for which experimental data is lacking for conventional VLE calculations, e.g. via Peng Robinson~\cite{robinson1985pr_eos}.
	
	\section{Conclusion}
	
	\noindent We incorporate the Clapeyron equation into the training of GNNs for predicting single-species vapor-liquid equilibria from molecular graph structure and temperature in a multi-task learning setting.
	We find Clapeyron regularization of model training to greatly increase the approximation of the exact Clapeyron equation while maintaining the same level of prediction accuracy.
	Additionally, multi-task learning of vapor molar volume, liquid molar volume, vapor pressure and enthalpy of vaporization significantly increases prediction accuracy compared to single-task learning. 
	Clapeyron-GNN is therefore a highly promising architecture for single-species VLE predictions for molecular and process design applications.
	
	Future work should investigate the possibility of using the Clapeyron equation as a hard constraint in the model architecture, following our recently introduced thermo\-dynamics-consistent GNN approach~\cite{rittig2024thermodynamics_consistent}, ie., embedding the Clapeyron equation directly into the output head of the model.
	This would be particularly interesting, as the present results, in particular for the enthalpy, show that employing physics-based regularization does not guarantee that all predictions are physically consistent in all scenarios.
	Notably, our first investigations of a thermodynamics-consistent GNN with embedded Clapeyron equation did result in lower prediction accuracy, requiring further tests to be performed in future work, targeting a wider range of properties and molecules. 
	For extending data sets and model testing in practical scenarios, collaboration with industry thus remains critical.
	
	\section*{CRediT authorship contribution statement}
	\textbf{Jan Pav\v{s}ek:} Conceptualization, Methodology, Software, Formal analysis, Investigation, Writing - original draft, Visualization.
	\textbf{Alexander Mitsos:} Conceptualization, Writing - review \& editing, Supervision,  Funding Acquisition.
	\textbf{Elvis J. Sim:} Methodology, Software, Investigation, Writing - review \& editing, Visualization.
	\textbf{Jan G. Rittig:}
	Conceptualization, Software, Writing - original draft, Supervision, Funding Acquisition.
	
	\section*{Data and Software Availability}
	
	\noindent The data used for training of the presented models are confidential and can be accessed through the NIST Thermodata engine. The code is available in our \emph{GitLab} repository \href{https://git.rwth-aachen.de/avt-svt/public/gmolprop/}{\emph{GMoLprop}}.
	
	\section*{Acknowledgments}
	\noindent This project was funded by the Deutsche Forschungsgemeinschaft (DFG, German Research Foundation) – 466417970 – within the Priority
	Programme ‘‘SPP 2331: Machine Learning in Chemical Engineering’’. 
	
	Model training was performed with computing resources granted by RWTH Aachen University.
	
	The authors thank René Görgen for his software engineering support.
	
	\bibliographystyle{elsarticle-num}  
	\bibliography{literature}

@software{nist_tde,
  author       = {{National Institute of Standards and Technology (NIST)}},
  title        = {ThermoData Engine, NIST Standard Reference Database 103b, Version 10.4.5},
  year         = {2025},
  version      = {Version 10.4.5 (2025)},  
  url          = {https://www.nist.gov/mml/acmd/trc/thermodata-engine/srd-nist-tde-103b},
  note         = {Accessed June 20, 2025; critically evaluated thermodynamic and transport property data},
}

@article{winter2025understanding,
  title={Understanding the language of molecules: predicting pure component parameters for the PC-SAFT equation of state from SMILES},
  author={Winter, Benedikt and Rehner, Philipp and Esper, Timm and Schilling, Johannes and Bardow, Andr{\'e}},
  journal={Digital Discovery},
  volume={4},
  number={5},
  pages={1142--1157},
  year={2025},
  publisher={Royal Society of Chemistry}
}

@article{rittig2024thermodynamics_consistent,
  title={Thermodynamics-consistent graph neural networks},
  author={Rittig, Jan G and Mitsos, Alexander},
  journal={Chemical Science},
  volume={15},
  number={44},
  pages={18504--18512},
  year={2024},
  publisher={Royal Society of Chemistry}
}

@article{specht2024hanna,
  title={HANNA: hard-constraint neural network for consistent activity coefficient prediction},
  author={Specht, Thomas and Nagda, Mayank and Fellenz, Sophie and Mandt, Stephan and Hasse, Hans and Jirasek, Fabian},
  journal={Chemical Science},
  volume={15},
  number={47},
  pages={19777--19786},
  year={2024},
  publisher={Royal Society of Chemistry}
}

@article{park2025vleGNN,
  title={A Multi-Stage Graph Neural Network--Physics-Informed Neural Network (GNN--PINN) Framework for Thermodynamic Property Prediction},
  author={Park, Jinyoung and Muthoka, Ruth M and Jang, Sunghyun and Lee, Yongjin},
  journal={Industrial \& Engineering Chemistry Research},
  volume={64},
  number={40},
  pages={19722--19734},
  year={2025},
  publisher={ACS Publications}
}

@article{schweidtmann2020graph,
  title={Graph neural networks for prediction of fuel ignition quality},
  author={Schweidtmann, Artur M and Rittig, Jan G and Konig, Andrea and Grohe, Martin and Mitsos, Alexander and Dahmen, Manuel},
  journal={Energy \& fuels},
  volume={34},
  number={9},
  pages={11395--11407},
  year={2020},
  publisher={ACS Publications}
}

@article{rittig2023gibbs_informed,
  title={Gibbs--Duhem-informed neural networks for binary activity coefficient prediction},
  author={Rittig, Jan G and Felton, Kobi C and Lapkin, Alexei A and Mitsos, Alexander},
  journal={Digital Discovery},
  volume={2},
  number={6},
  pages={1752--1767},
  year={2023},
  publisher={Royal Society of Chemistry}
}

@inproceedings{FeyLenssen2019pytorchgeom,
  title={Fast Graph Representation Learning with {PyTorch Geometric}},
  author={Fey, Matthias and Lenssen, Jan E.},
  booktitle={ICLR Workshop on Representation Learning on Graphs and Manifolds},
  year={2019},
}

@article{robinson1985pr_eos,
  title={The development of the Peng-Robinson equation and its application to phase equilibrium in a system containing methanol},
  author={Robinson, Donald B and Peng, Ding-Yu and Chung, Samuel YK},
  journal={Fluid Phase Equilibria},
  volume={24},
  number={1-2},
  pages={25--41},
  year={1985},
  publisher={Elsevier}
}

@article{kochi2025thermodynamics,
  title={Thermodynamics-informed machine learning for predicting temperature-dependent chemical properties},
  author={Kochi, Mahyar Rajabi and Rezaei, Hanie and Khan, Sartaaj Takrim and Mamillapalli, Bhanu Teja and Ebrahimiazar, Maryam and Ye, Haoming and Moosavian, Rose and Zargartalebi, Mohammad and Sinton, David and Moosavi, Seyed Mohamad},
  year={2025}
}

@article{heid2023chemprop,
  title={Chemprop: a machine learning package for chemical property prediction},
  author={Heid, Esther and Greenman, Kevin P and Chung, Yunsie and Li, Shih-Cheng and Graff, David E and Vermeire, Florence H and Wu, Haoyang and Green, William H and McGill, Charles J},
  journal={Journal of Chemical Information and Modeling},
  volume={64},
  number={1},
  pages={9--17},
  year={2023},
  publisher={ACS Publications}
}

@article{raissi2019physics,
  title={Physics-informed neural networks: A deep learning framework for solving forward and inverse problems involving nonlinear partial differential equations},
  author={Raissi, Maziar and Perdikaris, Paris and Karniadakis, George E},
  journal={Journal of Computational physics},
  volume={378},
  pages={686--707},
  year={2019},
  publisher={Elsevier}
}

@article{rittig2025molecular_perspective,
  title={Molecular machine learning in chemical process design},
  author={Rittig, Jan G and Dahmen, Manuel and Grohe, Martin and Schwaller, Philippe and Mitsos, Alexander},
  journal={arXiv preprint arXiv:2508.20527},
  year={2025}
}

@article{medina2026graph_margules,
  title={Graph neural networks embedded into Margules model for vapor--liquid equilibria prediction},
  author={Medina, Edgar Ivan Sanchez and Sundmacher, Kai},
  journal={Fluid Phase Equilibria},
  volume={599},
  pages={114520},
  year={2026},
  publisher={Elsevier}
}

@article{qin2023solvgnn,
  title={Capturing molecular interactions in graph neural networks: a case study in multi-component phase equilibrium},
  author={Qin, Shiyi and Jiang, Shengli and Li, Jianping and Balaprakash, Prasanna and Van Lehn, Reid C and Zavala, Victor M},
  journal={Digital Discovery},
  volume={2},
  number={1},
  pages={138--151},
  year={2023},
  publisher={Royal Society of Chemistry}
}

@article{wahyudi2026deepthermomix,
  title={DeepThermoMix: a local composition graph neural networks model for multicomponent activity coefficients},
  author={Wahyudi, Apri and Sueviriyapan, Natthapong and Rirksomboon, Thirasak and Suriyapraphadilok, Uthaiporn and others},
  year={2026}
}

@article{leenhouts2025thermodynamics_gnn,
  title={Thermodynamics-informed graph neural networks for phase transition enthalpies},
  author={Leenhouts, Roel and Jankelevitch, Sebastien and Raike, Roel and M{\"u}ller, Simon and Vermeire, Florence},
  journal={Systems and Control Transactions},
  pages={1662--1669},
  year={2025}
}

@article{bin2025exploring_augment,
  title={Exploring data augmentation: Multi-task methods for molecular property prediction},
  author={bin Javaid, Muhammad and Gervens, Timo and Mitsos, Alexander and Grohe, Martin and Rittig, Jan G},
  journal={Computers \& Chemical Engineering},
  pages={109253},
  year={2025},
  publisher={Elsevier}
}

@inproceedings{standley2020tasks,
  title={Which tasks should be learned together in multi-task learning?},
  author={Standley, Trevor and Zamir, Amir and Chen, Dawn and Guibas, Leonidas and Malik, Jitendra and Savarese, Silvio},
  booktitle={International conference on machine learning},
  pages={9120--9132},
  year={2020},
  organization={PMLR}
}

@article{rosenberger2022thermoconsistent,
  title={Machine learning of consistent thermodynamic models using automatic differentiation},
  author={Rosenberger, David and Barros, Kipton and Germann, Timothy C and Lubbers, Nicholas},
  journal={Physical Review E},
  volume={105},
  number={4},
  pages={045301},
  year={2022},
  publisher={APS}
}

@article{alam2026equinet,
  title={EquiNet: Modeling Vapor-Liquid Equilibrium Using Simultaneous Neural-Network Predictions for Activity Coefficient and Vapor Pressure},
  author={Alam, M Zaher and Alqallaf, Abdullah and McGill, Charles},
  year={2026},
  publisher={ChemRxiv}
}

\end{document}